\documentclass[conference]{IEEEtran}
\IEEEoverridecommandlockouts
\usepackage{cite}
\usepackage{comment}
\usepackage{balance}
\usepackage{amsmath,amssymb,amsfonts}
\usepackage{algorithmic}
\usepackage{graphicx}
\usepackage{textcomp}
\usepackage{xcolor}
\usepackage{tikz}
\usepackage{pgfplots}
\usepackage{hyperref}
\usepackage{multirow}
\usepackage{multicol}
\usepackage{blindtext, rotating}
\pgfplotsset{compat=newest}
\usepackage{tcolorbox}
\definecolor{mygr}{HTML}{e6e6e6}
\usepackage{listings}
\usepackage{xcolor}
\usepackage{tabularx,booktabs}
\definecolor{trans}{HTML}{FDAE61}
\definecolor{rand}{HTML}{ABDDA4}
\definecolor{original}{HTML}{2B83BA}

\definecolor{airforceblue}{rgb}{0.36, 0.54, 0.66}
\definecolor{brown(web)}{rgb}{0.65, 0.16, 0.16}
\definecolor{backcolour}{rgb}{0.95,0.95,0.92}
\definecolor{copper}{rgb}{0.72, 0.45, 0.2}
\definecolor{cobalt}{rgb}{0.0, 0.28, 0.67}
\definecolor{arsenic}{rgb}{0.23, 0.27, 0.29}
\definecolor{burlywood}{rgb}{0.87, 0.72, 0.53}
\definecolor{bronze}{rgb}{0.8, 0.5, 0.2}
\definecolor{bondiblue}{rgb}{0.0, 0.58, 0.71}
\lstdefinestyle{code_list}{
    keywordstyle=\color{blue},
    numberstyle=\tiny,%\color{},
    stringstyle=\color{brown(web)},
    basicstyle=\ttfamily\footnotesize,
    breakatwhitespace=false,         
    breaklines=true,                 
    captionpos=b,                    
    keepspaces=true,                 
    numbers=left,                    
    numbersep=5pt,                  
    showspaces=false,                
    showstringspaces=false,
    showtabs=false,                  
    tabsize=2,
    otherkeywords = { :,>=, <=, !=, True, False, ++},
    keywordstyle = [2]{\color{red}},
    morekeywords = [2]{},
    keywordstyle = [3]{\color{bronze}},
    morekeywords = [3]{factorial},
    keywordstyle = [4]{\color{cobalt}},
    morekeywords = [4]{1, 2, n,i, fact},
    keywordstyle = [5]{\color{bondiblue}},
    morekeywords = [5]{}
}

\newcommand{\defensepinku}[1]{{\color{black} #1}}
\newcommand{\icsmepinku}[1]{{\color{black} #1}}

\def\BibTeX{{\rm B\kern-.05em{\sc i\kern-.025em b}\kern-.08em
    T\kern-.1667em\lower.7ex\hbox{E}\kern-.125emX}}
\begin{document}

%\title{Demystifying the Current State of Deep Learning Models for Semantic Clone Detection }
\title{
On the Use of Deep Learning Models for Semantic Clone Detection
}

\author{
% Submission ID: 211 
\IEEEauthorblockN{Subroto Nag Pinku}
\IEEEauthorblockA{\textit{Department of Computer Science} \\
\textit{University of Saskatchewan}\\
Saskatoon, Canada \\
subroto.npi@usask.ca}
\and
\IEEEauthorblockN{Debajyoti Mondal}
\IEEEauthorblockA{\textit{Department of Computer Science} \\
\textit{University of Saskatchewan}\\
Saskatoon, Canada \\
d.mondal@usask.ca}
\and
\IEEEauthorblockN{Chanchal K. Roy}
\IEEEauthorblockA{\textit{Department of Computer Science} \\
\textit{University of Saskatchewan}\\
Saskatoon, Canada \\
chanchal.roy@usask.ca}
\and
}

\maketitle

\begin{abstract} 
Detecting and tracking code clones can ease various software development and maintenance tasks when changes in a code fragment should be propagated over all its copies. Several deep learning based clone detection models have appeared in the literature for detecting syntactic and semantic clones, and these models have widely been evaluated with the BigCloneBench dataset. However, the class imbalance and small number of semantic clones make BigCloneBench less ideal when interpreting the model performances. Sometimes researchers use a few other semantic clone datasets such as GoogleCodeJam, OJClone and SemanticCloneBench to understand a model's generalizability. To overcome the limitations of the existing datasets, recently a GPT-assisted large semantic and cross-language clone dataset GPTCloneBench has been released, but it is not clear how all these models would compare and contrast in terms of these datasets. In this paper, we propose a multi-step evaluation approach for five state-of-the-art clone detection models leveraging existing benchmark datasets including the recently proposed GPTCloneBench and exploiting the mutation operators to study the extent of these clone detection models' ability. More specifically, we examined the performance of three highly-performing single-language clone detection models (ASTNN, GMN, CodeBERT) that use various code representations (e.g., AST, flow augmented AST with graph matching network, and bidirectional encoder representation) for detecting semantic clones. In addition to using BigCloneBench, we tested them on SemanticCloneBench and GPTCloneBench, investigated their robustness under mutation operations, and examined them against cutting-edge cross-language clone detection tools (C4, CLCDSA) that are also known to learn semantic clones. While all single-language models showed high F1 scores for BigCloneBench, their performances varied quite differently (sometimes over 20\%) when tested on SemanticCloneBench. Interestingly, the cross-language model (C4)  consistently showed superior performance (around 7\%) on SemanticCloneBench over other models and performed similarly for BigCloneBench and GPTCloneBench. On mutation-based datasets, C4 appeared to have a more robust performance (less than 1\% difference) whereas single-language models showed high variability. 
\end{abstract} 


\section{Introduction}
Duplicate code fragments in software systems are known as code clones. Software systems often have around 9\% to 17\% code clones in them\cite{zibran2011analyzing, baxter1997software, baker1995finding}. Developers introduce clones in software systems during development by reusing existing code, copying-pasting codes, etc. Clones incur a considerable cost on software maintenance tasks\cite{mondal2017does}. A rich body of research is now available on clone detection~\cite{wang2023comparison,farmahinifarahani2019precision,DBLP:journals/jss/MondalRS20,roy2007survey}, visualization~\cite{DBLP:journals/vi/MondalMRSLW19,DBLP:conf/icse/MondalMRSWL19,hammad2020systematic,DBLP:conf/scam/UddinGGR15}, and analysis~\cite{DBLP:books/sp/21/IR2021,nguyen2011clone,dang2017transferring,kim2005empirical,DBLP:journals/jss/NadimMRS22} of code clones both in single-language~\cite{DBLP:journals/jss/NasrabadiPRRE23,roy2007survey} and cross-language settings~\cite{mehrotra2023improving,DBLP:conf/iwsc/PinkuMR23,DBLP:conf/iwsc/RoyAARRS23}. In this paper we exclusively focus on the clone detection models, in particular those that are built using deep learning.

Several deep learning based clone detection models have appeared in the literature for detecting syntactic and semantic clones, where semantic clones, i.e., codes with the same functionality but different syntax, are believed to be more challenging to detect, and even more so in a cross-language setting. The approaches for single-language code clone detection\cite{zhang2019novel, wang2020detecting, feng2020codebert} have been very successful in existing evaluations for syntactic clone detection \cite{lei2022deep}. However, detecting clones with the same semantics but different syntax is challenging~\cite{DBLP:conf/iwpc/KeivanlooRR12,pinku2023pathways,wu2020scdetector}. Figure \ref{fig:examplsemanctic} depicts an example of semantic clones. These code fragments calculate the factorial of a given number, but the logic behind their computation is different. The differences in syntax and lexemes are easily noticeable.

The state-of-the-art clone detection techniques are based on deep learning\cite{lei2022deep} and evaluating them can play a big role in being used in the real world\cite{alam2023gptclonebench, svajlenko2021bigclonebench, al2020semanticclonebench, krinke2022bigclonebench, nafi2019clcdsa, pinku2023pathways}. \icsmepinku{As a result,} many benchmark datasets have been used in the literature for such evaluation\cite{bellon2007comparison, svajlenko2021bigclonebench, al2020semanticclonebench, alam2023gptclonebench}. Earlier benchmark datasets such as Bellon's work in \cite{bellon2007comparison} were built through manual validation of the contemporary subject systems but relied on tools that are now outdated. %As a result,
\icsmepinku{Consequently, majority of the }modern single-language clone detection models are evaluated on the BigCloneBench dataset\cite{svajlenko2021bigclonebench,DBLP:conf/iwsc/SvajlenkoR22}. This dataset consists of clone and non-clone pairs based on structural similarity\icsmepinku{, and it} does not provide any code pairs that are entirely semantic clones, i.e., structurally very different but functionally similar. Moreover, the definition of semantic clones is ambiguous \icsmepinku{and often debated in literature as it}  %and it
revolves around syntactic similarity\cite{svajlenko2021bigclonebench, blass2009two, yernaux2022detecting}. %Consequently, 
\icsmepinku{Thus,} evaluating semantic clone detection models on BigCloneBench %(which was 
(originally created for measuring recall\cite{svajlenko2021bigclonebench}) may result in misleading evaluation. Furthermore, BigCloneBench has class imbalance issues due to the chosen design principle\cite{krinke2022bigclonebench, alam2023gptclonebench}, whereas machine learning models are often built and compared on a balanced dataset. \icsmepinku{Therefore, BigCloneBench may not be well suited for developing and evaluating machine-learning based clone detection models\cite{krinke2022bigclonebench, yu2022assessing}}. %reference-added

A few other datasets \icsmepinku{created} from programming contests such as GoogleCodeJam(GCJ)\cite{wang2020detecting}, OJClone\cite{wei2017supervised}
\cite{mou2016convolutional} are sometimes used to evaluate single-language clone detection models%' performances
\cite{lei2022deep}. To alleviate the lack of availability for semantic clones, Al-Omari\cite{al2020semanticclonebench} et al. proposed SemanticCloneBench that consists of single-language semantic clone pairs for \icsmepinku{ four programming %different 
languages i.e., Java, Python, C\#, and C++.} However, this dataset only contains 1,000 pairs for each language and is often insufficient to build a robust machine-learning model.

\icsmepinku{By the definition, } clones across different languages (cross-language) are also semantic clones due to the inherent property of the languages' syntax being very different from one to another\cite{pinku2023pathways}. There are only a handful of deep learning-based techniques available for cross-language clone detection\cite{vislavski2018licca, cheng2017clcminer, tao2022c4, nafi2019clcdsa, pinku2023pathways}. Cross-language datasets are scarce \icsmepinku{in literature \cite{pinku2023pathways, nafi2019clcdsa}.} \icsmepinku{Though SemanticCloneBench\cite{al2020semanticclonebench} has data for four programming languages, it does not provide any cross-language clone data.} The CLCDSA dataset\cite{nafi2019clcdsa} 
is the only widely known dataset for cross-language clones created by curating programming competition data. To overcome some of the limitations of the existing datasets (e.g., size, types of clones, imbalance issues, number of available programming languages, lack of validation), Alam et al, recently proposed a GPT-assisted semantic and cross-language clone dataset, GPTCloneBench with  37,149 single-language semantic clone pairs and 20,770 cross-language clone pairs. These clones were validated in multiple ways including human validation, functionality testing, and tool-assisted validation \cite{alam2023gptclonebench}.

\begin{figure}[t!]
\begin{lstlisting}[xleftmargin=0.5cm,columns=flexible,language=Java,numbers=none, style=code_list]
public long factorial(int n) {
    long fact = 1;
    for (int i = 2; i <= n; i++) {
        fact = fact * i;
    }
    return fact;
}
\end{lstlisting}
\textbf{\footnotesize (a) Java code to calculate factorial using loop} %\newline
\begin{lstlisting}[xleftmargin=0.5cm,columns=flexible,language=Java, numbers=none, style=code_list]
public long factorial(int n){
    if (n <= 2) {
        return n;
    }
    return n * factorial(n - 1);
}
\end{lstlisting}
 \textbf{\footnotesize (b) Java code to calculate factorial using recursion}\newline
\begin{lstlisting}[xleftmargin=0.5cm,columns=flexible,language=Python, numbers=none, style=code_list]
def factorial(n):
   if n == 1:
       return n
   else:
       return n*factorial(n-1) 
\end{lstlisting}
\textbf{\footnotesize (c) Python code to calculate factorial using recursion}
\caption{Example of semantic code clones. These code fragments implement the same function. The code in (a) and (b) are single-language clones. Code in (c) is a cross-language clone with (a) and (b).
}
\label{fig:examplsemanctic}
\end{figure}

\icsmepinku{The lack of consistent evaluation of semantic clone detection models\cite{alam2023gptclonebench} and the limitation of earlier evaluations based on BigCloneBench motivated us to examine the current benchmarking of clone detection models.}  We aim to reveal the performance variations among different models while utilizing diverse code representations and benchmark datasets. We chose five cutting-edge clone detection models that incorporate both single-language and cross-language approaches. These models include, ASTNN\cite{zhang2019novel}, GMN\cite{wang2020detecting}, CodeBERT\cite{feng2020codebert}, CLCDSA\cite{nafi2019clcdsa}, and C4\cite{tao2022c4}. They incorporate a variety of code representations, as well as distinct network architectures and learning objectives. For instance, the C4 model is built on the pre-trained CodeBERT and employs contrastive learning techniques \cite{frosst2019analyzing}, whereas, the CLCDSA model leverages hand-crafted features and a \icsmepinku{deep} siamese network. More specifically, we are interested in investigating how these models understand the different code representations (i.e., AST, Graph, Token, Features) for semantic clones. 

The reason we bring cross-language clone detection models into the scene is their strong ability to recognize code semantics across different languages, which naturally makes them comparable against single-language clone detection tools when targeting semantic clones. \icsmepinku{This also enables us to see how different training methods can impact clone detection across benchmarks. }

Existing clone detection tools are sometimes evaluated based on mutation and injection frameworks \cite{roy2009mutation} that generate synthetic clones through different mutation operations. Therefore, after examining these models' behaviour for standard benchmark datasets, we %also wanted to see how they behave for mutation. 
\icsmepinku{are also interested in examining the performance behaviour of these models when they are presented with a change in the code fragments through mutation.} 
\defensepinku{A model’s robustness is often examined based on its behaviour under an adversarial input which is obtained by tweaking the  input\cite{henkel2022semantic} through various  transformations\cite{yu2022data}. Since these transformations are similar to mutations, we leverage mutation-based modification to the input to test a model's robustness. Although a mutation can generate artificial syntactic clones that may not preserve the exact original semantics, the pair arguably would still have similar representations for the models to detect them as clones.}

\smallskip
\noindent
In particular, we consider the following research questions.

\smallskip
\noindent
\textbf{RQ1. \textit{How do the cutting-edge single-language clone detection models compare with each other when detecting semantic clones?}}

\smallskip
\noindent
To answer this research question, we chose three state-of-the-art single-language clone detection techniques that includes CodeBERT\cite{feng2020codebert}, ASTNN\cite{zhang2019novel}, and GMN\cite{wang2020detecting}. We intend to find the capabilities of these models in the context of semantic code clones. We evaluated these models using SemanticCloneBench\cite{al2020semanticclonebench} and GPTCloneBench\cite{alam2023gptclonebench}, as well as with BigCloneBench\cite{svajlenko2021bigclonebench} dataset to understand their relative performances.\smallskip

\smallskip
\noindent
\textbf{RQ2. \textit{How do the cross-language clone detection models compare with single-language clone detection models for detecting syntactic or semantic clones? }}

\smallskip
\noindent
In this research question, we explored the potential of cross-language clone detectors to detect single-language semantic clones, which seems natural as cross-language models learn representation by training through datasets where the code fragments have little syntactic similarities. We evaluated the baseline, CLCDSA and state-of-the-art model C4 by leveraging the CLCDSA dataset\cite{nafi2019clcdsa}, BigCloneBench\cite{svajlenko2021bigclonebench}, SemanticCloneBench\cite{al2020semanticclonebench}, and GPTCloneBench dataset.\smallskip

\smallskip
\noindent
\textbf{RQ3. \textit{How robust are these models when evaluated with semantic clones that are mutated with common mutation operations?}}

\smallskip
\noindent
To answer this research question, we used the mutation-based modifications on the SemanticCloneBench dataset. Then we evaluated the models' performance for each type of modification. We analyzed the average performance for each type of operator, and how the performance varied with respect to the performance on the original SemanticCloneBench dataset.

\smallskip
\noindent
\textbf{Our contribution.} In this paper, we answer the research questions RQ1--RQ3  and make the following contributions.

We conduct extensive experiments and show that traditional training and testing of deep learning based clone detection models using one or two datasets may not reveal their true performance, whereas a multi-step evaluation can provide better insights into the models' potential. We evaluate three single-language clone detection models\cite{wang2020detecting, feng2020codebert, zhang2019novel} with BigCloneBench\cite{svajlenko2021bigclonebench}, SemanticCloneBench \cite{al2020semanticclonebench}, and GPTCloneBench\cite{alam2023gptclonebench}. Our experimental results show that the performance of single-language models varies differently up to 20\% when tested on a semantic clone dataset. 

To understand how the cross-language clone detection tools compare with the single-language ones we evaluate them with both BigCloneBench\cite{svajlenko2021bigclonebench}, SemanticCloneBench\cite{al2020semanticclonebench}, GPTCloneBench\cite{alam2023gptclonebench} along with CLCDSA dataset\cite{nafi2019clcdsa}. The experimental results show that the cross-language model, C4, has a superior performance of up to 7\%. We exploited the mutation operators\cite{svajlenko2019mutation, roy2009mutation} to make modifications to the semantic clone dataset as a means to test the clone detection models' robustness. The results show that different models perform differently depending on the type of mutation operator. The performance does not vary more than 1\% for the cross-language model C4, which appears to be more robust for mutation changes than other models.

\section{Related Work} 

In this section we discuss the related research. We first discuss the clone detection approaches, then the datasets that have been used for evaluation, and finally, the work related to cross-language clone detection. 

\subsection{Code Clone Detection}
Clone detection techniques can broadly be divided into single-language and cross-language clones\cite{roy2007survey, lei2022deep}.

\subsubsection{Single Language}
There exist many techniques that can detect syntactic clones with high accuracy such as the text and transformation based\cite{roy2008nicad}, token-based \cite{li2004cp}, PDG-based\cite {higo2011incremental} or Tree-based\cite{koschke2006clone} techniques. Several recently proposed clone detection methods are based on deep learning. ASTNN, a novel neural representation of source code, employs a tree-based approach to classify source codes and detect clones\cite{zhang2019novel}. Wang et al.\cite{wang2020detecting} proposed a graph matching network that detects clones using a graph neural network architecture. They used control/data flow information with graph representation of source code and evaluated the model using BigCloneBench and GoogleCodeJam. In another work,  Wei et al. proposed the CDLH framework which learns the deep features using AST-based LSTM\cite{wei2017supervised}. Deepsim\cite{zhao2018deepsim} achieved better results in detecting similar functionalities and surpassed many other techniques such as CDLH\cite{wei2017supervised}, RtvNN\cite{white2016deep}, and DECKARD\cite{jiang2007deckard}. They extracted semantic features of code fragments to train a deep neural network. Feng et al.\cite{feng2020codebert}  recently developed CodeBERT, a transformer-based neural architecture, that appeared to be very effective in generating code representations for %deep learning based 
clone detection. 

\subsubsection{Cross Language}

% Code fragments that are written in two different languages and have similar functionalities are considered semantic clones in a cross-language setting. 
\icsmepinku{There are only a small number of techniques available for semantic clone detection in cross-language settings\cite{lei2022deep}.} Nafi et al.\cite{nafi2019clcdsa} designed a deep Siamese network-based cross-language clone detector. They extracted 9 features from the code fragments to train the deep learning model. Mathew et al. proposed a language agnostic code clone detection technique \cite{mathew2020slacc}, as well as a code search technique based on static and dynamic analyses which could be used as a special case for cross-language clone detection\cite{mathew2021cross}. Among the other techniques, CLCMiner\cite{cheng2017clcminer} requires code revision histories from the code repository, and LICCA\cite{vislavski2018licca} depends on a tree-based approach combined with a modified version of the longest common subsequence algorithm to detect cross-language clones. %C4 is 
A recent technique, C4, based on pre-trained CodeBERT achieved state-of-the-art performance\cite{tao2022c4}. %. It exploited contrastive learning to build and train the model.

\subsection{Datasets and Benchmarking}
Around 90\% of studies in the literature involve clone detection in single-language settings and the rest involve cross-language clone detection\cite{lei2022deep}. We now give a brief overview of such clone detection tools and the datasets that were used to evaluate them.

\subsubsection{Evaluation of Clone Detection Tools with Benchmark Datasets}

Bellon et al. evaluated six clone detectors on large software systems written in C and Java \cite{bellon2007comparison}. They built a reference corpus to benchmark by manually validating the clones detected for further evaluation. Their selected tools were able to detect syntactical clones and the representations used were Token, AST, PDG, and function metrics. Previous studies\cite{murakami2014dataset, svajlenko2014evaluating} showed that this type of evaluation might not be accurate. For this reason, the mutation and injection framework is often used to evaluate clone detection tools\cite{svajlenko2019mutation}. However, this framework generates artificial clones. To test clone detection \icsmepinku{tools' performance} %models
on real-world software data, Svajlenko et al. created a benchmark dataset\cite{svajlenko2021bigclonebench}. We refer the readers to \cite{bellon2007comparison, svajlenko2019mutation, svajlenko2014evaluating, svajlenko2015evaluating, lei2022deep} for early works on comparing clone detection models which are mostly focused on syntactic clones. Our work differs from that of Svajlenko et al.\cite{svajlenko2015evaluating, svajlenko2014evaluating} as our target deep learning techniques are capable of detecting semantic clones. 

\icsmepinku{A few recent works have investigated deep learning models' generalization ability and performance on semantic clone detection\cite{choi2023investigating, arshad2022codebert, rabbani2022comparative} including the different interpretations of codes by humans and models through explainable AI techniques (i.e., SHAP)\cite{abid2023interpreting}. We take a different approach by introducing a multi-level evaluation of available benchmarks and exploiting mutation operations to assess robustness.} %reference-added

\subsubsection{Single Language}
Svajlenko et al. created the BigCloneBench dataset from known clones of the IJadataset\cite{svajlenko2021bigclonebench} source repository to use as a benchmark dataset to evaluate clone detection tools and techniques. Other datasets such as OJClone\cite{zhang2019novel} and GoogleCodeJam dataset\cite{wang2020detecting} are created from programming competition data and are sometimes used to evaluate single-language clone detection tools as well. However, these code fragments \icsmepinku{from programming competition} are not from real-world software. Code fragments in these datasets might contain different ranges of syntactic similarity among them. To resolve these issues, a crowd-sourced dataset, SemanticCloneBench, was proposed by Al-Omari et al.\cite{al2020semanticclonebench}.
%It contains 4,000 semantic clone pairs for four programming languages. i.e., Java, Python, C\#, and C++. The authors also provided software systems with clone pairs injected into them. This dataset was created by filtering Stack-Overflow posts to create 1,000 clone pairs for each of four programming languages  
\icsmepinku{It only provides clone-pairs and} does not provide any non-clone pairs which are essential for a balanced dataset to train a deep learning based classifier\cite{al2020semanticclonebench}. Additionally, a standalone clone pairs dataset was provided in SemanticCloneBench that can be used to %build and 
evaluate clone detection techniques and tools. %However, this dataset appears to be small 
\icsmepinku{However, the number of pairs might not be considered sufficient }% from the perspective of 
training deep learning based clone detection models.

\icsmepinku{With the advancements in Generative AI\cite{vaswani2017attention, radford2018improving, radford2019language}}, Alam et al. proposed GPTCloneBench, a dataset that leveraged the SemanticCloneBench and GPT model to create semantic clone pairs\cite{alam2023gptclonebench}. Through prompting the GPT with code fragments from SemanticCloneBench, the authors created a large number of clone pairs and then filtered syntactic clones using a syntactic clone detector NiCAD\cite{roy2008nicad}. This dataset considers syntactic clone pairs \icsmepinku{(see Section \ref{sec: clone-cat})} %(Type-I to Type-III)
as negative samples. The primary goal of providing such pairs as negative samples is to provide a balanced dataset that enables a machine learning model to learn to differentiate between semantic clones and non-semantic clones\cite{alam2023gptclonebench}. However, such negative samples are not non-clone pairs which is different from the negative samples that are commonly created for training deep learning based clone detection. 

\subsubsection{Cross Language}
CLCDSA dataset\cite{nafi2019clcdsa} is commonly being used to evaluate cross-language clone detection models\cite{mathew2021cross, tao2022c4}. This dataset provides code fragments in five different languages. However, several studies used a filtered version of this dataset, e.g., using only a subset of the provided code fragments such as those written in Java and Python\cite{pinku2023pathways, mathew2021cross}. %Several researchers
\icsmepinku{A number of researches produced} %used their own 
dataset to evaluate their model\cite{vislavski2018licca, cheng2017clcminer}. The GPTCloneBench dataset also provides %true 
clone pairs for cross-language clones and does not include any negative samples\cite{alam2023gptclonebench}.

To the best of our knowledge, none of the earlier evaluation techniques focused on multi-step evaluation of deep learning models such as ours through syntactic and semantic benchmarks along with mutation data \icsmepinku{through different representation and training methods (i.e., single-language and cross-language)}.

\section{Background}

In this section, we discuss the necessary background.

\subsection{Clone Detection}
\label{sec: clone-cat}
Code clones are often categorized into four types based on their syntactic and functional nature\cite{roy2007survey}.
\subsubsection{Type-I}
Code fragments are similar in syntax and only differ in white spaces and comments.
\subsubsection{Type-II}
This type of clone has changes in their identifier names and literal values along with Type-I changes.
\subsubsection{Type-III}
The difference in statement level is defined as Type-III clones. These statements are added, removed or modified. 
\subsubsection{Type-IV}
Code fragments that have similar functionality but are syntactically dissimilar are known as Type-IV clones. For example, bubble sort and quick sort, both sort algorithm has the same functionality but have different syntax. Type-IV clones are known as semantic clones, and cross-language clones also belong to this category.

\subsection{Source Code Representation}
The models we studied in this work require the following code representations. 

\subsubsection{Abstract Syntax Tree (AST)}
Abstract Syntax Tree is the syntactical graph-based representation of code fragments. Each node represents a construct (i.e., variable types, names, statements, conditionals) in the written program or formal language. They hold the necessary syntactical information and preserve the flow of control of the code fragments as well. An AST is a directed graph starting from the root node (e.g., the main function), and ending with leaf nodes (e.g., identifiers).

\subsubsection{Flow Augmented Abstract Syntax Tree (FA-AST )}
Wang et al.\cite{wang2020detecting} modified AST with 11 additional edges to provide more control and data flow information. This empowers ASTs with more semantic information about the code. but adding more edges also increases the computational and structural complexity of the graph.

\subsubsection{Textual Representation (Token)}
Techniques from natural language processing (NLP) are often adopted for source code analysis because of the similarity in texts' nature. \icsmepinku{
One of the techniques involves representing codes as tokens.} Tokens are defined as the smallest unit of a statement, such as words. This representation usually requires a dictionary that maps each token to a number, and by using it, a word embedding is learned at some point in the algorithm, \icsmepinku{capturing semantic information and relationships between tokens.}

\subsection{Techniques chosen for this study}
In the literature, many clone detection techniques have been proposed that work well for Type-I to Type-III clones. Several of these methods show high accuracy also for detecting Type-IV clones when using existing benchmark datasets. We focus on the state-of-the-art deep learning models for code clone detection, \icsmepinku{selecting them based on their reported performance in the literature and their focus on semantic clones\cite{zhang2019novel, feng2020codebert, wang2020detecting, tao2022c4, nafi2019clcdsa}.} %which we select based on their performances reported in the literature and whether they target the problem of detecting semantic clones.

\subsection{Single-Language Clone Detection Models}

\begin{comment}
\begin{figure}
  \centering
  \includegraphics[keepaspectratio, width=0.35\textwidth]{graph-fast.drawio.png}
    \caption{AST(blac edges) and FA-AST(red edges) representation}
    \label{fig:qf}
\end{figure}
\subsubsection{Flow Augmented Abstract Syntax Tree with Graph Neural Network}
\end{comment}

\subsubsection{Graph Matching Network (GMN)}
Wang et al. \cite{wang2020detecting} used FA-AST representation with graph matching network to build a clone detection model. The graph matching network\cite{li2019graph} takes two graphs and jointly learns the embedding. Then, it calculates cross-graph attention among nodes of the given graphs. As a consequence, the GMN model requires graphs to be homogeneous (i.e., code fragments from the same programming language). Once it learns embedding for the nodes, it uses a readout function to get graph-level embedding. These two graph-level embeddings are then used to calculate similarity.

\subsubsection{AST-based Neural Network (ASTNN)}
Zhang et al. proposed an AST-based neural source code representation\cite{zhang2019novel} for code classification and code clone detection. The authors first split the ASTs into statement levels and then created multi-way statement trees. After that, a recursive neural network was used to learn the vector representation of the statements trees. The word2vec\cite{word2vec} model was used to embed the code fragments\cite{zhang2019novel}. For an AST $T$,  statement trees $t$, are extracted and encoded as vector $Q \in \mathbf{R}^{T \times k} = [e_1, e_2,...,e_T], t \in [1, T]$ and $k$ is the encoding dimensiion. Once the embedding for each statement tree is learned, the rest of the framework is defined as standard bidirectional GRU %as proposed in 
\cite{bahdanau2014neural}. With the help of these encoding techniques and dynamic batching, the representations are learned in this algorithm\cite{zhang2019novel}.

 %%%%%%%%
\begin{figure*}[!t]
  \centering
  \includegraphics[keepaspectratio, width=.6\textwidth]{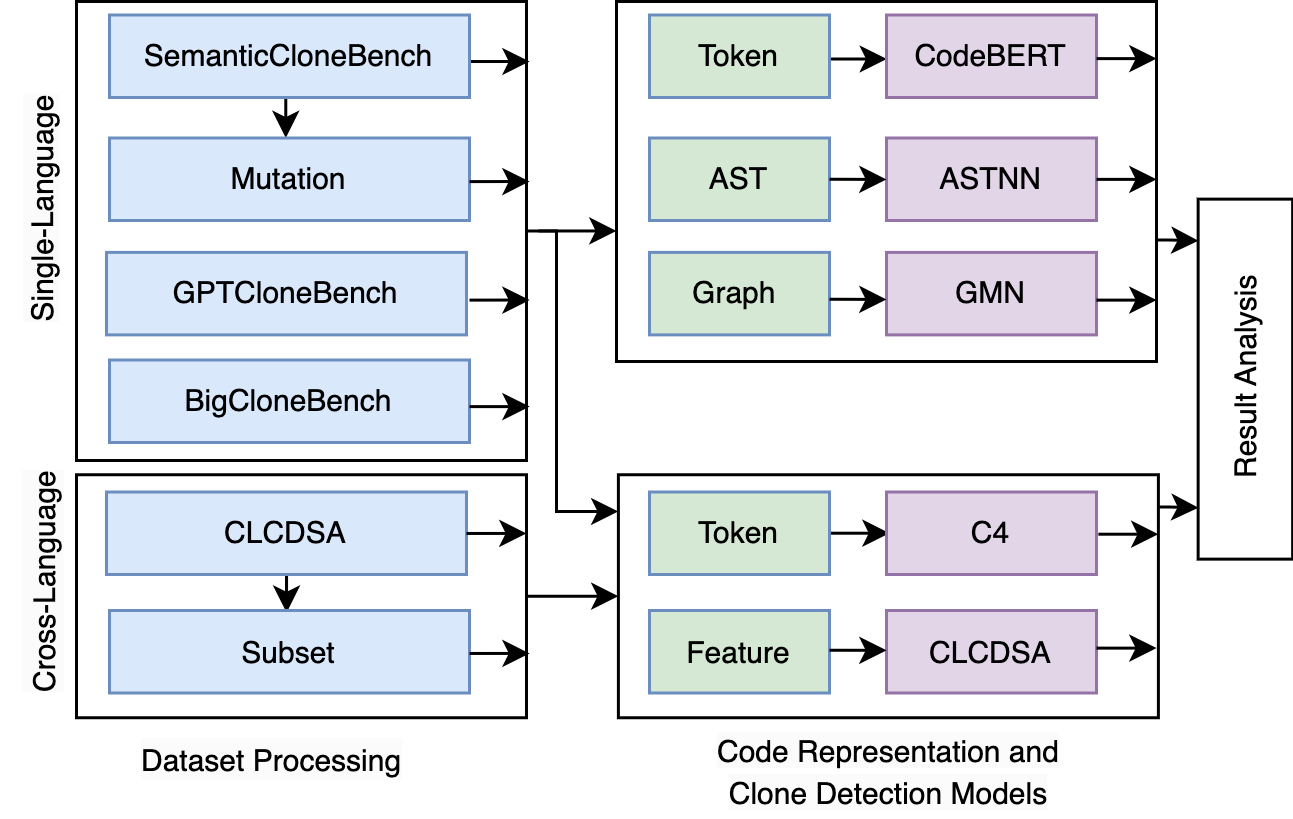}
    \caption{\icsmepinku{Overview of our approach. The left side shows the dataset processing. The right side shows the input representations and clone detection models.}}
    \label{fig:Overview}
\end{figure*}

%%%%%%%%%%

\subsubsection{Pre-Trained Model for Programming and Natural Language}

CodeBERT\cite{feng2020codebert} is a pre-trained model for natural language (NL) and programming language (PL) tasks such as code search, clone detection, and code summarization. It is derived from BERT\cite{devlin-etal-2019-bert} and the architecture it follows is the same as RoBERTa\cite{roberta}. For training, it concatenates string pairs as  $[CLS]w_1, w_2, ... , w_n[SEP], c_1, c_2, ..., c_n [EOS]$, where $CLS$, $SEP$ and $EOS$ represent special tokens such as classification,  sentence separator, and end of sentence respectively. \icsmepinku{Words and codes are denoted using $w$ and $c$, and this string pair is known as NL-PL pair as per the order of appearance of natural language (NL) and programming language (PL).} %The CLS denotes that it is a classification problem. 

The same idea can be extended to PL-PL tasks, such as given two code fragments, the model can tell us about their similarity. CodeBERT is adopted for any specific task by further training (fine-tuning) the model with domain-specific data.

\subsection{Cross-Language Clone Detection Models}
 \subsubsection{CLCDSA}
 The CLCDSA is a widely used model for cross-language clone detection\cite{nafi2019clcdsa}. It is developed on siamese architecture of the deep neural network and requires pre-computed features.  These features are derived from the Software Quality Observatory for Open Source Software (SQO-OSS) quality model\cite{samoladas2008sqo, software-quality-monden}. The dataset used in this paper is commonly used among researchers to evaluate cross-language clone detection techniques and it's the only well-known dataset for this task\cite{nafi2019clcdsa, tao2022c4}. This model has been used as a baseline for \icsmepinku{the development of many other} models as well\cite{tao2022c4, mathew2021cross}.
 
 \subsubsection{C4}
 C4 is the state-of-the-art cross-language clone detection model built leveraging contrastive-learning loss function with pre-trained CodeBERT\cite{tao2022c4}. The overall framework for contrastive learning involves a couple of steps\cite{frosst2019analyzing}. First, each code fragment is required to be presented as a vector. Second, positive and negative examples are constructed for each sample vector. Finally, the distance between these pairs is calculated using a similarity function to train the model using contrastive learning. 
 
\subsection{Mutation-based analysis for code clone}
Mutation is the process of modifying code fragments through a series of defined atomic operations primarily used in software testing\cite{bradbury2007comparative}. It has also been used in code clone analysis. In particular, mutation and injection-based frameworks have been used to evaluate code clone detection tools\cite{svajlenko2019mutation} and primarily used to evaluate the recall of real-world software systems. The standard atomic operations are defined in literature\cite{roy2009mutation} which can be exploited further to modify code fragments to test clone detection techniques\cite{svajlenko2019mutation}.

%old figure

\section{Methodology}
This section describes the methodology we followed.

\subsection{Problem Formulation}
We define the clone detection problem as follows:
\icsmepinku{Two code fragments \(C_i\) and \(C_j\) with a label \(y_{ij}\) are considered a clone pair if their syntactic or semantic similarity exceeds a specified threshold; otherwise, they are classified as a non-clone pair. The label $y_{ij}$ %denotes if it's a clone-pair or non-clone pair.
indicates whether the pair is a clone or non-clone. Let $s_{ij}$ be a score that indicates the similarity between the given pair of code fragments.} %Given a threshold $\sigma$, a true clone pair is determined based on whether $s_{ij}> \sigma$ or not.  
\icsmepinku{For any threshold $\sigma$, a pair is considered clone if $s_{ij}> \sigma$, otherwise they are not clone.}

\subsection{Dataset Selection and Processing}

\defensepinku{We used three single-language clone datasets and one cross-language clone dataset for this study. In particular, the three single langaue benchmarks include BigCloneBench\cite{svajlenko2021bigclonebench}, SemanticCloneBench\cite{al2020semanticclonebench}, and GPTCloneBench\cite{alam2023gptclonebench} and for cross-langauge, we used CLCDSA\cite{nafi2019clcdsa} dataset. The SemanticCloneBench considers answers for the same question from Stack-Overflow as clone pairs, we carefully shuffled them to create non-clone pairs so that none of the non-clone pairs contained answers from the same Stack-Overflow question\cite{wang2020detecting}. Additionally, to understand how the models perform on a dataset that can distinguish semantic clones from the rest of the clone pairs, we used the GPTCloneBench\cite{alam2023gptclonebench}.  In this study, we utilized code fragments written in Java from the datasets GPTCloneBench\cite{alam2023gptclonebench} and SemanticCloneBench\cite{al2020semanticclonebench}. We did not take any datasets from the programming competition as real-world benchmark datasets are available for single-language clones. Moreover, single-language clone detection models have already performed very high on real-life benchmark datasets\cite{wang2020detecting, zhang2019novel} and thus those competitive programming datasets are often not used for model evaluation in single-language settings. Table \ref{table:single} shows the statistics for BCB, SCB, and GPTCB that represent BigCloneBench, SemanticCloneBench and GPTCloneBench respectively. Positive samples are clone pairs for all three datasets and negative samples are non-clone pairs for BCB and SCB. \icsmepinku{However, GPTCB only has all positive clone pairs but that paper has also provided semantic pairs (Type-IV) and non-semantic pairs (Type-I to Type-III, i.e., negative pairs in Table-\ref{table:single}). Therefore, here models are separating Type-IV from the rest (unlike classifying clone vs non-clone pairs).}

%%%%% single lang datasets table
 \begin{table}[t]
\caption{Summaries of BigCloneBench (BCB), SemanticCLoneBench (SCB) and GPTCloneBench (GPTCB) Dataset}
\label{table:single}
\centering
\begin{tabular}{ l r  r r}
\hline
 %&\multicolumn{2}{c}{ } &\multicolumn{2}{c}{}\\

Metric &BCB &SCB &GPTCB\\
\hline
\#Code fragments &8,070 &1,228 & 11,655\\
\#Average lines of codes  &33.21 &17.44 &16.73 \\
\#Positive Pairs  &561,929 &997  &5,711 \\
\#Negative Pairs &1,172,451  &993  &5,431 \\ 
\hline
\end{tabular} 
 \end{table}

%%%% cross lang data table
 \begin{table}[h]
\caption{Summaries of CLCDSA and CLCDSA$\_{sub}$ Dataset}
\label{table:cross}
\centering
\begin{tabular}{ l r  r }
\hline
 %&\multicolumn{2}{c}{ } &\multicolumn{2}{c}{}\\

Metric &CLCDSA &CLCDSA$\_{sub}$\\
\hline
\#Programming language &5 &2 \\
\#Code fragments &83,560 &38,167 \\
\#Average lines of codes &157.22 &67.81\\
\#Clone pairs &138,804 & 312,582  \\
%\#False clone pairs &138,804 &156,291  \\
\hline

\end{tabular} 
 \end{table}

%%%%%%%%%%%%%%%%%

For CLCDSA, we used the version of the dataset provided in the replication package of the original paper %which is 
used in C4\cite{tao2022c4, nafi2019clcdsa}. C4 only considers one random code fragment from each class of problem. As a result, the number of pairs is not very large. Since several studies that develop cross-language models have created their dataset\cite{vislavski2018licca} or used a subset of the CLCDSA dataset\cite{mathew2021cross} it leads to an inconsistent performance comparison. To mitigate this problem, we filtered a subset of the CLCDSA dataset that includes code fragments from Java and Python programming languages. Such filtering has also been done in the literature\cite{mathew2021cross, mathew2020slacc, pinku2023pathways}. We refer to this as CLCDSA$_{sub}$. The statistics for CLCDSA and CLCDSA$_{sub}$ are given in Table \ref{table:cross}.}

\subsection{Mutation-based Dataset Generation}
In addition to using the standard datasets from literature\cite{nafi2019clcdsa, svajlenko2021bigclonebench, al2020semanticclonebench, alam2023gptclonebench}, we also create a mutation-based dataset that consists of modified code fragments to evaluate the robustness of the models. We used a standard set of mutation operations that consists of  15 atomic operations \cite{svajlenko2019mutation}. \defensepinku{Svajlenko et al. provided details of these operators in their paper\cite{svajlenko2019mutation}, and we leveraged the mutation operators from the framework\footnote{https://github.com/jeffsvajlenko/MutationInjectionFramework} for our purpose.} \icsmepinku{We applied each mutation operator on SCB to get 15 new datasets. In this step, these mutated datasets and the original SCB remain unseen during training and models are tested against them (Table \ref{table:mutation-all}). The results on mutated datasets provide us with an insight into the model's generalization ability, robustness, and performance\cite{zhang2023challenging} through the lens of the performance variation on each type of mutation, which in turn ensures that the models are not only accurate but also resilient to code changes. We also compare type-wise average performance with the original SCB (Table \ref{table:mutation-avg}). By analyzing performance across various mutation types, we can identify specific weaknesses and strengths in the models. This holistic approach provides a deeper understanding of how well the models can handle real-world scenarios where code evolves.}

\subsection{Code Representation and Feature Extraction}
We aim to study different state-of-the-art techniques, and the representations of code fragments are different for each algorithm. We created different representations for different techniques. This representation includes tokens, abstract syntax tree  (AST), and flow-augmented abstract syntax tree (FA-AST). Moreover, embedding plays a pivotal role in deep-learning algorithms since it facilitates the features to be represented as numbers for the algorithms. The token-based techniques could directly take the code fragments as input with a little preprocessing such as tokenization, and removing comments. In the case of CodeBERT, tokenization was done by the pre-trained tokenizer provided with CodeBERT\cite{feng2020codebert}.  

For the graph-based techniques\cite{wang2020detecting}, the nodes were embedded by taking the word embedding of a node type and textual representation\cite{allamanis2018learning}. ASTNN uses one-hot-encoding and a bidirectional GRU for the learning task\cite{zhang2019novel}. For the cross-language settings, the CLCDSA model requires %relies on pre-computed 
features extracted from the AST representation of the code fragments through the ANTLR parser\footnote{https://www.antlr.org/}\cite{nafi2019clcdsa}. The C4 model is built %leveraging contrastive learning with 
on top of CodeBERT thus the input is the same as the CodeBERT model\cite{tao2022c4}.

\subsection{Evaluation Metrics}
All models are evaluated using precision, recall, and F1 score following their original implementations\cite{wang2020detecting, zhang2019novel, feng2020codebert, nafi2019clcdsa, tao2022c4}. Precision captures the effect of large negative samples on classifiers though it is sensitive to data distribution. However, recall is not distribution sensitive and does not tell the number of incorrectly classified positive samples. The F1 score is used as the ultimate measure that combines both precision and recall. They are defined as follows.
\icsmepinku{
\begin{equation}
P = \dfrac{C_r}{C_r+I_c}; \text{ }
\end{equation}

\begin{equation}
R = \dfrac{C_r}{C_r+ I_{nc}} ;\text{ } 
\end{equation}

\begin{equation}
F1 = \dfrac{2 \times P \times R}{P + R}
\end{equation}
}
% \begin{equation}
%     Precision = \dfrac{TP}{TP+FP}
% \end{equation}
% \begin{equation}
%     Recall = \dfrac{TP}{TP+ FN}    
% \end{equation}

In the above equations, $C_r$ is the number of positive samples retrieved correctly, $I_c$ denotes the number of negative samples incorrectly classified as positive, and the number of positive samples that were mistaken by the model as negative is denoted by $I_{nc}$. 

\subsection{Experimental Settings}
We followed the experimental setting described in the respective papers\cite{wang2020detecting, zhang2019novel, feng2020codebert, nafi2019clcdsa, tao2022c4}. For single-language models, we used the dataset and followed the pair creation procedure as described by Wang et al.\cite{wang2020detecting, svajlenko2021bigclonebench, al2020semanticclonebench, alam2023gptclonebench}. For cross-language models, we used the CLCDSA dataset\cite{nafi2019clcdsa} and pairs given in the replication packages of respective models\cite{nafi2019clcdsa, tao2022c4} and the subset was created by filtering Java and Python code fragments\cite{nafi2019clcdsa}.  We split the datasets 8:1:1 for training, testing and validation data. All models were trained on a machine using RTX-3080Ti. Our codes and datasets are available \href{https://github.com/subrotonpi/clone_evaluation}{here\footnote{https://github.com/subrotonpi/clone\_evaluation}}. 
%--------------------------
\section{Evaluation and Discussion}

In this section, we describe the result of our experimental evaluation and answer the research questions.

\smallskip
\noindent
\textbf{RQ1. \textit{How do the cutting-edge single-language clone detection models compare with each other when detecting semantic clones?}}

We trained and tested single-language models on BigCloneBench, SemanticCloneBench, and GPTCloneBench to understand their performances for semantic clone detection. Table \ref{table: bcb} shows that the models achieved almost the same F1 score ranging between 93\% to 95\% on the BigCloneBench dataset. The graph-matching network has the highest precision, and CodeBERT has the highest recall value among the techniques. 

\begin{table}[h]
\caption{Results when trained and tested on BigCloneBench}

\begin{center}
\begin{tabular}{l l c c c} 
 \hline
 Code Representation & Model & Precision & Recall & F1 \\
 \hline 
AST& ASTNN&	0.938&	0.936&	0.937\\
\hline 
FA-AST &GMN &\textbf{0.968}&	0.931&	\textbf{0.949}\\
% & GGNN&	0.338&	\textbf{0.987}&	0.504\\
\hline 
Token& CodeBERT& 0.927&	\textbf{0.946}&	0.936\\
\hline
 \end{tabular}
\end{center}
\label{table: bcb}
 \end{table}

%-------
  \begin{table}[h]
\caption{Results when trained on BigCloneBench and Tested on SemanticCloneBench}

\begin{center}

\begin{tabular}{l l c c c} 
 \hline
 Code Representation & Model & Precision & Recall & F1 \\
 \hline 
 AST & ASTNN &0.658	&0.389	&0.489\\
 \hline
FA-AST & GMN &0.540	&\textbf{0.918}	&\textbf{0.680} \\
 \hline
 % \multirow{1}{4em}{
 Token & CodeBERT &\textbf{ 0.697}	&0.466	&0.559\\
\hline
\end{tabular}
\end{center}
\label{table: bcb-scb}
 \end{table}

Many studies use an alternative dataset to measure the performance of deep learning models on unseen data\cite{yu2022data, nafi2019clcdsa}. As BigCloneBench is a large dataset, we investigated how the model trained using it performs on the SemanticCloneBench dataset. Table \ref{table: bcb-scb} shows the evaluation of the SemanticCloneBench dataset. Additionally, the BigCloneBench and SemanticCloneBench datasets are from real-world software systems and forums, respectively. Hence one may expect the models to perform quite similarly. However, we found that there is a significant drop in the F1 score ranging from 27\% for GMN to 47\% for ASTNN. Even in this case, the GMN model performed well with the highest recall and F1 score. While CodeBERT appears to be better in terms of precision,  GMN shows more stability when considering recall values.

%--------

 \begin{table}[h]
\caption{Results when trained and tested on SemanticCloneBench}

\begin{center}

\begin{tabular}{l l c c c} 
 \hline
 Code Representation & Model & Precision & Recall & F1 \\
 \hline 
 AST & ASTNN & 0.882 & \textbf{0.891} & 0.886\\
 \hline
 FA-AST 
& GMN & \textbf{0.944} & 0.841 & \textbf{0.890} \\
% & GGNN & 0.716 &0.950 & 0.817 \\
 \hline
 \multirow{1}{4em}{Token} 
%& CodeBERT & 0.75 & 0.48 & 0.59\\
& CodeBERT$_{scb}$ & 0.789 & \textbf{0.891} & 0.837\\
\hline
\end{tabular}
\end{center}
\label{table: scb}
 \end{table}
%--------

 \begin{table}[h]
\caption{Results when trained and tested on GPTCloneBench}

\begin{center}

\begin{tabular}{l l c c c} 
 \hline
 Code Representation & Model & Precision & Recall & F1 \\
 \hline 
 AST & ASTNN & 0.894 & 0.908 & 0.901\\
 \hline
 FA-AST 
& GMN & 0.787 & 0.906 & 0.842 \\
 \hline
 \multirow{1}{4em}{Token} 
& CodeBERT$_{gptcb}$ &\textbf{ 0.996 }& \textbf{0.983} & \textbf{0.990}\\
\hline
\end{tabular}
\end{center}
\label{table: gptcb}
 \end{table}

Table \ref{table: scb} and table \ref{table: gptcb} show the results of the single-language models when trained and tested on the SCB, and GPTCB datasets respectively. We observe a drop in the models' performances on SCB when compared to the performances on BCB (Table \ref{table: bcb}). However, CodeBERT achieved around 83\% F1 score and its performance increased by about 24\% when fine-tuned with SCB. Also, in this case, GMN performed better than all others, though the current baseline model for single-language clone detection, ASTNN performed similarly and has the highest recall value. Table \ref{table: gptcb} shows that on GPTCB the fine-tuned CodeBERT achieved the best performance while ASTNN performed similarly. However, the GMN model's performance dropped by around 5\% than on SCB. 

%  \begin{table}[h]
% \begin{center}
%     \caption{Occurrences of Control Flow Nodes}
%     % for FA-AST}% in SemanticCloneBench} 
%     %24876 4024 3851 29899 130 109
%     %gpt 32508 4089 17301 52840 331 668
    
%     \begin{tabular}{l r r r}
%         \hline
%         Control Structure & SCB &BCB &GPTCB\\
%         \hline
%         If-else & 2,896 &24,876 & 32,508\\
%         %\hline
%         While  & 310 &4,024 & 4,089\\
%         %\hline
%         For & 1,596 &3,851 &17,301 \\
%         %\hline
%         Block  & 4,649 &29,899 &52,840\\
%         %\hline
%         Do-while & 26 &130 & 331\\
%         %\hline
%         Switch-case & 87 &109 & 668\\
%         \hline
% \end{tabular}
%  \label{table:faastnodes}
% \end{center}
%  \end{table}

\subsection*{Choosing a single-language clone detection model.}  
\noindent
Traditional evaluation under BCB shows that the three models vary less than 1\% in F1 scores. However, the GMN has the highest F1 score and the effectiveness of the graph matching network comes from its ability to capture the structure better than other representations. The graph-based model seems to generalize well to distinguish clone pairs from non-clone pairs in BCB and SCB (\ref{table: bcb}, \ref{table: bcb-scb}, \ref{table: scb}. When \icsmepinku{trained on BCB and tested} with SCB, the GMN model shows a 12\% higher F1 score than CodeBERT (Table \ref{table: bcb-scb}). It achieved 6\% more F1 score compared to CodeBERT even when fine-tuned with SemanticCloneBench (Table \ref{table: scb}). The GMN model has around 20\% better performance than ASTNN (Table \ref{table: bcb-scb}). However, ASTNN has almost similar performance (less than 1\% difference) as GMN when trained with SCB (Table \ref{table: scb}). 

\icsmepinku{The GMN model tends to fail in distinguishing Type-IV clones from other types of clones in GPTCB.
% In the case of GPTCB, the GMN model did not perform well.  
The probable reasons for this are that GPTCB has insignificant structural similarity and the dataset contains only clone-pairs which might make it difficult for a graph-based model that relies primarily on structural similarity %the pairs are clone pairs in the samples 
\cite{alam2023gptclonebench} }

 \begin{table}[h]
\begin{center}
    \caption{Number of trainable parameters in clone detection  models}% in thousands
    \begin{tabular}{l l r}
        \hline
         Type & Model &\#Parameters\\
        \hline
        \multirow{3}{4em}{Single-language} 
        &ASTNN &138 K\\%138,458 \\
         &GMN &123 K\\%\\%122,801\\ %7,856,201 
         % &GGNN &103 K\\%102,701\\
         &CodeBERT &125 M\\%125,827K\\%125,827,586
        \hline
        \multirow{2}{4em}{Cross-language}
        &CLCDSA &47 K\\ %47,891
         & C4 &172 M \\ %172, 503 K %172, 503, 552
        \hline
\end{tabular}
 \label{table:parameters}
\end{center}
 \end{table}

If we consider the number of trainable parameters in models \defensepinku{(Table \ref{table:parameters})}, the graph matching network (GMN) is a reasonable choice for single-language clone detection. Although models like ASTNN have an almost similar number of parameters, the performance of GMN makes it a preferred choice; keeping in mind that the GMN model performed better when providing non-clone pairs as opposed to distinguishing among different clone types\cite{alam2023gptclonebench}. Being a pre-trained large language model, CodeBERT has way more parameters which are on a scale of millions. Though, it does not achieve the best result for semantic clones as seen in Table \ref{table: scb}, it could well learn to distinguish among clone types in the GPTCB dataset\cite{alam2023gptclonebench}. As GPTCB is made by prompting GPT with SCB code fragments, it seems to provide data that follows certain distribution\cite{tang2023science} and fine-tuning CodeBERT with a large number of pairs helped it to learn the dataset better as opposed to less number of fragments in SCB.
\smallskip

\begin{tcolorbox}[left=0pt,right=0pt,top=1pt,bottom=1pt,boxrule=0.1pt,colback=black!5!white,colframe=black!75!black]
In summary, we found that the nature of the datasets largely impacts the models' performance even for large models. When detecting semantic clones, the graph-based techniques show more robustness and better performance compared to the baseline and state-of-the-art models up to 20\% in terms of F1 score. However, CodeBERT performs better in separating semantic clones from other types of clones. 
\end{tcolorbox}

%Note to self for r2/3 ?: 

% We did not test on SCB after training with BCB, the reason is that we want to see if cross-language clone detection models learn representation well enough through cross-lang training. Since it is more semantic than single lang datasets like BCB.

% This is the same for testing on Mutation data. We did not consider BCB since it contains highly syntactically similar fragments. We are interested in scenarios with the lowest syntactic similarity.

% We did not train ASTNN, and GMN with the cross-language dataset. Because, ASTNN learns representation for a single language, and GMN requires homogenous graphs. In the case of CodeBERT, it could be used but it does not support all programming languages in the dataset.

\smallskip
\noindent
\textbf{RQ2. \textit{How do the cross-language clone detection models compare with single-language clone detection models for detecting syntactic or semantic clones? }}

\icsmepinku{So far, }we have explored how different single-language clone detection models perform in detecting semantic clones. Since cross-language clone detectors are generally meant to be language agnostic\cite{mathew2020slacc}, we now examine how they compare with single-language clone detectors.

%%%%%%%%%%%%%%%
%----cross-clcdsa
\begin{table}[h]
\caption{Cross-Language clone detection results on CLCDSA dataset}
\begin{center}
\begin{tabular}{l c c c c} 
\hline 
Code Representation &Model& Precision &Recall &F1\\
\hline
Extracted Features &CLCDSA& 0.49& 0.83& 0.62\\
\hline
Token &C4& \textbf{0.920}& \textbf{0.913}& \textbf{0.916}\\
\hline
\end{tabular}
\label{table:clcdsadata}
\end{center}
 \end{table}
%%%%%%%%%%%%%%%

We trained the cross-language models on the CLCDSA dataset
\cite{tao2022c4, nafi2019clcdsa}. Since these models used different numbers of languages in their experiments,  we created a subset of the dataset\cite{mathew2020slacc, mathew2021cross} by choosing Java and Python to compare them on the same ground and to see how their performance varies. Table \ref{table:clcdsadata} shows the performance of the models on the CLCDSA dataset. We see that the baseline, CLCDSA model has poor precision but a high recall value. The C4 model has both high precision and recall values. For the subset data, Table \ref{table:clcdsa-sub-data} shows that the overall performance of CLCDSA increased but its precision is still the same, whereas C4 achieved around a 5\% increase in performance. 
%%%%%%%%%%%%%%%
%---- cross-clcdsa-sub
\begin{table}[h]
\caption{Cross-Language clone detection results on CLCDSA$_{sub}$}
\begin{center}
\begin{tabular}{l c c c c c} 
\hline 
Code Representation &Model& Precision &Recall &F1\\
\hline
Extracted Features &CLCDSA& 0.49& 0.99& 0.66\\
\hline
Token &C4& \textbf{0.95}& \textbf{0.96}& \textbf{0.96}\\
\hline
\end{tabular}
\label{table:clcdsa-sub-data}
\end{center}
 \end{table}
 %%%%%%%%%%%%
 
  %---- cross-bcb
 \begin{table}[htbp]
\caption{Cross-Language models' performance trained and tested on BigCloneBench}
\begin{center}
\begin{tabular}{l c c c c c} 
\hline 
Code Representation &Model& Precision &Recall &F1\\
\hline
Extracted Features &CLCDSA& 0.58  & 0.96  & 0.72  \\
\hline
Token &C4  &\textbf{0.920} &\textbf{0.946 } &\textbf{0.933}\\
\hline
\end{tabular}
\label{table:cross-bcb-data}
\end{center}
 \end{table}
 
Now that we have an understanding of these models' performance for cross-language clones, %which are semantically the same,
we want to see their performances on single-language benchmark datasets. Table \ref{table:cross-bcb-data} shows the model's performance on the BCB dataset. We can see that the relative performance is better since the code fragments in BCB have high syntactical similarity whereas the CLCDSA dataset's codes are written by individuals from scratch. More interestingly, the performance of C4 is similar to the single language clone detection models that were trained and tested with BCB (Table \ref{table: bcb}).

%---- cross-clcdsa-scb
\begin{table}[h]
\caption{Cross-Language models' performance trained on CLCDSA and tested on SemanticCloneBench}
\begin{center}
\begin{tabular}{l c c c c c} 
\hline 
Code Representation &Model& Precision &Recall &F1\\
\hline
Extracted Features &CLCDSA &0.414	&0.164	&0.234 \\
\hline
Token &C4 &\textbf{0.613}	&\textbf{0.713	}&\textbf{0.659} \\
\hline
\end{tabular}
\label{table:cross-train-scb-test}
\end{center}
 \end{table}
 %----

 %--- cross scb
\begin{table}[h]
\caption{Cross-Language models' performance trained and tested on SemanticCloneBench}
\begin{center}
\begin{tabular}{c c c c c} 
\hline 
Code Representation &Model& Precision &Recall &F1\\
\hline
Extracted Features &CLCDSA &0.48	&0.65	&0.55 \\
\hline
Token &C4 &\textbf{0.953}	&\textbf{0.975} &\textbf{0.963} \\
\hline
\end{tabular}
\label{table:cross-scb}
\end{center}
 \end{table}
 %----

   %---- cross-gptcb
 \begin{table}[tp]
\caption{Cross-Language models' performance trained and tested on GPTCloneBench}
\begin{center}
\begin{tabular}{l c c c c c} 
\hline 
Code Representation &Model& Precision &Recall &F1\\
\hline
Extracted Features &CLCDSA & 0.510  & 0.877  & 0.645  \\
\hline
Token &C4  &\textbf{0.999} &\textbf{0.994} &\textbf{0.996} \\

\hline
\end{tabular}
\label{table:cross-gptcb-data}
\end{center}
 \end{table}

%------
%%%%%%%%%

Cross-language code fragments have semantic similarity but little syntactic similarity as opposed to single-language. We thus examine whether cross-language clone detection models learn representation well enough through cross-language training to detect clones in SCB. Table \ref{table:cross-train-scb-test} shows that the C4 model trained on the CLCDSA dataset and tested on SCB achieved nearly the performance of the single-language model, GMN's performance which was trained with the single-language clone dataset (Table \ref{table: bcb-scb}). This shows that the C4 model learned representation better than other models through the cross-language semantic data. When these models were trained and tested on the SCB dataset, we still noticed a decline in performance for the CLCDSA model. Table \ref{table:cross-scb} shows the results on SCB where we see that the C4 model still has a high F1 score of up to 7\% when compared to the best-performing single-language model, GMN (Table \ref{table: scb}). Table \ref{table:cross-gptcb-data} shows the results for CLCDSA and C4 models on the GPTCB dataset. The C4 model has also performed highly distinguishing Type-IV single language clone pairs. 

%%%%%%%%
\begin{figure}[h]
  \centering
  \includegraphics[keepaspectratio, width=.5\textwidth]{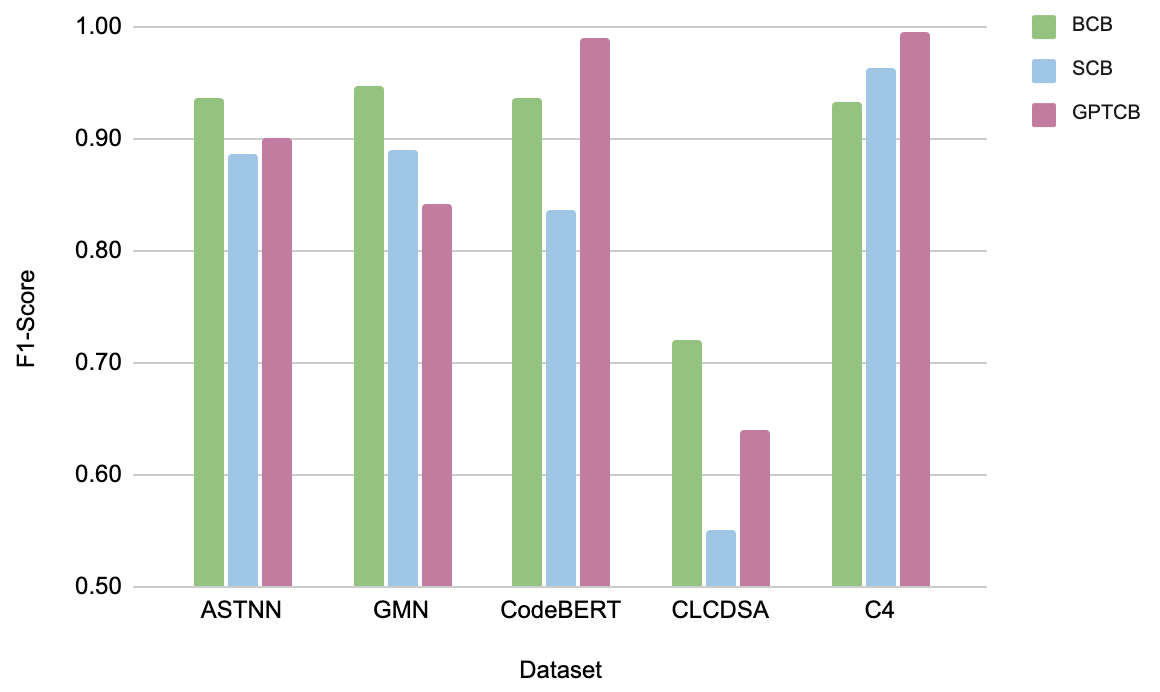}
    \caption{\icsmepinku{Models' performance when trained and tested on benchmark datasets}}
    \label{fig:models-vs-data}
\end{figure}

%%%%%%%%%%
To understand the relative difference better, Figure \ref{fig:models-vs-data} shows different models' performance across benchmark datasets. We see that single-language models were mostly consistent apart from that GMN's performance dropped on GPTCB. The C4 model mostly performed higher than other models on all three benchmark datasets whereas the CLCDSA consistently had lower performance. Though the CLCDSA model is widely used as baseline\cite{nafi2019clcdsa, mathew2021cross, tao2022c4}, in our evaluation, it performed poorly with respect to other models (around 30\%) and should be considered as a weak baseline\footnote{https://blog.ml.cmu.edu/2020/08/31/3-baselines/}. As a natural consequence, we believe, that a better choice of baseline is required in further research for developing a language-agnostic clone detection model. 

Initially, we thought that the cross-language clone detectors that learn representation in cross-language settings might not be as efficient as single-language models to detect semantic clones in single-language settings. However, our experiment shows the C4 model performed better in both cross-language clones and single-language semantic clone detection. Notice that C4 is computationally expensive as it is built on top of CodeBERT (\ref{table:parameters}). 
 
\begin{tcolorbox}[left=0pt,right=0pt,top=1pt,bottom=1pt,boxrule=0.1pt,colback=black!5!white,colframe=black!75!black]
In summary, our experimental results show that when detecting syntactic clones, the cross-language clone detection model C4 performs similarly to the single-language clone detection models that are trained on single-language clone datasets. However, when detecting semantic clones, C4  performs better than other models (sometimes up to 7\%).
\end{tcolorbox}

%----
 %---
%mutation all table
\begin{table*}
\caption{Performance of models on each mutation operation on SemanticCloneBench dataset}
\begin{center}
\setlength{\tabcolsep}{7pt}
\begin{tabular}{l l | c c c |   c c c |   c c c |   c c c |   c c c}
% |   c c c} 
\hline
\multicolumn{2}{c}{ } %empty 
&\multicolumn{3}{c}{ASTNN} 
&\multicolumn{3}{c}{GMN} 
% &\multicolumn{3}{c}{GGNN}
&\multicolumn{3}{c}{CodeBERT} 
&\multicolumn{3}{c}{CLCDSA}
&\multicolumn{3}{c}{C4}
\\
\hline %$\downarrow$ %$\rightarrow$
Types &Mutations  &P &R &F1 &P &R &F1 &P &R &F1 &P &R &F1 &P &R &F1 \\
% &P &R &F1 \\
\hline

\multirow{6}{*}{Type-I}
 &mCC\_BT 
&0.66	&0.39	&0.49 %astnn
&0.54	&0.93	&0.69 %gmn
% &0.50	&0.998	&0.67 %ggnn

&0.69	&0.44	&0.54 %codebert

&0.41	&0.16	&0.23 %clcdsa
&0.60	&0.72	&0.66 %c4
\\
&mCC\_EOL
&0.66	&0.39	&0.49 
&0.53	&0.91	&0.67 
% &0.50	&0.999	&0.67

&0.71	&0.46	&0.56

&0.41	&0.16	&0.23
&0.59	&0.71	&0.65 \\

&mCF\_A 
&0.66	&0.39	&0.49 
&0.53	&0.92	&0.68
% &0.50	&0.998	&0.67

&0.70	&0.47	&0.56 

&0.41	&0.16	&0.23
&0.60	&0.72	&0.66 \\

&mCF\_R 
&0.66	&0.39	&0.49 
&0.53	&0.92	&0.67
% &0.50	&0.999	&0.67

&0.69	&0.47	&0.56

&0.41	&0.16	&0.23
&0.62	&0.71	&0.66 \\

&mCW\_A
&0.66	&0.39	&0.49 
&0.53	&0.92	&0.67
% &0.50	&0.999	&0.67

&0.69	&0.46	&0.55

&0.41	&0.16	&0.23
&0.61	&0.71	&0.65 \\

&mCW\_R 
&0.66	&0.39	&0.49 
&0.53	&0.92	&0.68 
% &0.50	&0.999	&0.67

&0.70	&0.48	&0.57

&0.41	&0.16	&0.23
&0.60	&0.71	&0.65 \\
\hline

\multirow{4}{*}{Type-II}
&mARI  
&0.65	&0.38	&0.48 
&0.59	&0.76	&0.66
% &0.50	&0.993	&0.67

&0.70	&0.48	&0.57

&0.41	&0.16	&0.23
&0.61	&0.71	&0.66\\

&mRL\_N 
&0.71	&0.60	&0.65 
&0.63	&0.86	&0.73
% &0.50	&0.996	&0.67 

&0.71	&0.59	&0.65

&0.32	&0.10	&0.15
&0.62	&0.72	&0.67 \\

&mRL\_S 
&0.76	&0.63	&0.69 
&0.55	&0.95	&0.69
% &0.50	&1	&0.67

&0.73	&0.63	&0.68

&0.32	&0.12	&0.17
&0.55	&0.90	&0.68\\

&mSRI  
&0.64 &0.40	&0.49 
&0.55	&0.76	&0.64
% &0.50	&0.997	&0.67

&0.68	&0.47	&0.55

&0.42	&0.16	&0.23
&0.62	&0.71	&0.67 \\
\hline

\multirow{5}{*}{Type-III} 
&mDL  
&0.69	&0.43 &0.53 
&0.53	&0.88	&0.66
% &0.50	&0.999	&0.67

&0.70	&0.48	&0.57

&0.43	&0.17	&0.24
&0.61	&0.72	&0.66 \\

&mIL  
&0.82	&0.66 &0.73 
&0.53	&0.95	&0.68
% &0.59	&0.75	&0.67

&0.79	&0.62	&0.69

&0.34	&0.14	&0.19
&0.61	&0.72	&0.66\\

&mML  
&0.68	&0.45 &0.54 
&0.59	&0.75	&0.66
% &0.50	&0.995	&0.67

&0.71	&0.48	&0.57

&0.41	&0.10	&0.17
&0.60	&0.71	&0.65 \\

&mSDL  
&0.67 &0.41 &0.51 
&0.54	&0.77	&0.63
% &0.51	&0.877	&0.64

&0.69	&0.48	&0.57

&0.40	&0.15	&0.22
&0.60	&0.71	&0.65\\

&mSIL  
&0.67 &0.34	&0.45 
&0.57	&0.56	&0.56
% &0.52	&0.678	&0.59

&0.70	&0.47	&0.56

&0.43	&0.18	&0.25
&0.60	&0.70	&0.65 \\
\hline

\end{tabular}
\label{table:mutation-all}
\end{center}
 \end{table*}
%---

\smallskip
\noindent
\textbf{RQ3. \textit{How robust are these models when evaluated with semantic clones that are mutated with common mutation operations?}}

%So far 
\icsmepinku{Till this point,} we have leveraged existing standard datasets such as BCB, SCB, GPTCB, and CLCDSA datasets to understand the models' ability on semantic clone detection. One natural question to consider now is how these models would behave with altered inputs. \icsmepinku{To examine this, we equip us by exploiting }%we exploited 
the mutation operators to change the code fragments in the SCB dataset\cite{al2020semanticclonebench}. \defensepinku{The models are trained on the respective benchmark datasets\cite{svajlenko2021bigclonebench, nafi2019clcdsa} \icsmepinku{which enable us} to see how their performance varies based on different training approaches (i.e., single-language or cross-language) and against the mutant code fragments.} Table \ref{table:mutation-all} shows that each model performed similarly for all types of modifications for Type-I mutant fragments. Their performances decrease for modifications with Type-III mutation operations such as insertion, deletion or modification of a whole line.

We define robustness as a model's ability to perform under such situations. Table \ref{table:mutation-avg} shows the average F1 score for all models \icsmepinku{on the modified versions (mutants) }%on a mutant version 
of the SCB dataset. We notice that the models' performance varies mostly for Type-III mutations. It often decreases the performances as low as 4\% compared to the original SCB  in terms of F1 score. Among the single-language models, the GMN model works the best and has the lowest variation considering all types of mutations under these constraints. However, for Type-III mutations, where the mutation impacts a code fragment's syntax more, GMN's performance drops but ASTNN and CodeBERT appear to be more robust. This is because of ASTNN's capability to split the AST into subtrees whereas GMN uses single modified AST. For CodeBERT it uses tokens, and the similarity score tends to remain the same with original fragments. 

For cross-language models, Table \ref{table:mutation-avg} shows that C4 has the best performance across all types of modifications. Since C4 is built on top of CodeBERT with contrastive learning, it seems to be more robust when compared to other models. On the contrary, the CLCDSA model performed poorly when tested on the original SCB along with different types of mutation. \icsmepinku{This highlights the importance of leveraging advanced learning techniques to enhance model performance and robustness in various scenarios.}

%overall-mutation-avg
\begin{table}[t]
\caption{Overall performance on mutation-based datasets. Delta shows the difference between the models' performance on SemanticCloneBench (\ref{table: scb} \ref{table:cross-scb}) and the average F1 score for different types of mutation operations}
\begin{center}
\begin{tabular}{l|c r | c r  | c r } 
\hline
Types &\multicolumn{2}{c}{Type-I} &\multicolumn{2}{c}{Type-II}  &\multicolumn{2}{c}{Type-III}  \\
\hline
 Models  &F1$_{avg}$ & $\Delta_{SCB}$ &F1$_{avg}$ & $\Delta_{SCB}$ &F1$_{avg}$ & $\Delta_{SCB}$\\
\hline
ASTNN &0.489 &0\% &0.578 & 8.9\% &0.553 &6.4\% \\ 
GMN &0.677 &-0.3\% &0.682 &0.1\% &0.64 &-4\%\\ 
% GGNN &0.668 &0.02\% &0.668 &0\% &0.648 &-1.9\% \\ 
CodeBERT &0.557 &-0.2\% &0.613 &5.4\% &0.593 &3.4\% \\ 
\hline
CLCDSA &0.234 &0\% &0.235 &0.1\% &0.208 &-2.6\% \\ 
C4  &0.655 & -0.4\% &0.669 &0.01\% &0.655 &-0.4\% \\
\hline
\end{tabular}
\label{table:mutation-avg}
\end{center}
 \end{table}

\begin{tcolorbox}[left=0pt,right=0pt,top=1pt,bottom=1pt,boxrule=0.1pt,colback=black!5!white,colframe=black!75!black]
In summary, our experiments on mutation data show that models built on top of transformer-based pre-trained models are robust across common code mutations. The difference for various mutations is less than 1\% for C4 and can range between 3\% to 9\% for others.  This shows that cross-language training is more robust for semantic clone detection. 
\end{tcolorbox}

\section{Implications of our study}

Our experiments show that code representation greatly influences semantic clone detection. We also noticed that rudimentary representations, such as Tokens and Abstract Syntax Trees, are less likely to capture all aspects of a code's property. Along with FA-AST representation, graph-based techniques worked better for single-language clones. However, embedding through pre-trained models (i.e., CodeBERT) worked best \icsmepinku{in overall performance measure} since it could embed tokens more appropriately.  Practitioners can follow this evaluation path to further evaluate the deep learning-based clone detection tools and techniques rigorously. Considering the different needs, one could choose a pre-trained model if the goal is to locate clones with minimal effort, otherwise, other models can be used when there is sufficient time and resources to train a model. However, our experiment shows that C4 appears to be a high-performing and robust deep learning based option for semantic clone detection. 

Our findings suggest that the evaluation is largely dataset-dependent. As we have seen, LLM-based models (CodeBERT, C4) perform highly on LLM-generated datasets (GPTCB) and not very well in real-world stack overflow data (SCB). Though the intent of the two is not identical as GPTCB does not deal with non-clone pairs. Researchers can shed further light on exploring the different dataset properties causing the variation that might lead to better explainability of the models.

\section{Threats to validity}
In our work, we used the BigCloneBench\cite{svajlenko2021bigclonebench}, SemanticCloneBench\cite{al2020semanticclonebench}, GPTCloneBench\cite{alam2023gptclonebench} and the CLCDSA datasets. The CLCDSA dataset\cite{nafi2019clcdsa} consists of problems from two different programming competitions, and they are not from real-world software. However, these problems are more complex and require thoughtful coding to implement them. As a result, they give a proper ground to train and test a model's ability. The GPTCloneBench is also not from real-world software data. Nevertheless, it provides a very good ground to test models' performance on codes with very little syntactic similarity and the ability to distinguish Type-IV clones from the rest. \icsmepinku{Moreover, generative models are pre-trained on real-world data, and they are increasingly being adapted to create real-world codes.} 

Different clone detection models used different embedding techniques and standard algorithms, so the limitations applicable to them are also applicable to this work. These methods are generally well established and have also been used in prior works\cite{allamanis2018learning}. 

The SemanticCloneBench dataset has a small number of pairs for the classification task which might cause overfitting. However, cutting-edge models are tested against these cases and should be able to generalize well. Additionally, large language models are meant to be fine-tuned with small data for this kind of downstream task. The code granularity level is completely dependent on the model design which is usually method-level but a complete program could also be taken for some of the models. This is a design decision made by the respective authors of the different datasets and models. For single-language clone detection models, the Java programming language was used as it is widely used for different studies and the availability of benchmark datasets and model support. 

\defensepinku{We found that mutation does not change the models' performance to a large extent. This shows mutation does not change the code's behavior significantly.  Moreover, models learn the inherent semantics through representation learning of a code which is invariant to these types of changes. Some cases may have a normalization effect that increases the performance.} 

For cross-language datasets, we chose Java and Python to narrow down and find a way to have similar ground for model testing as they are the most used programming languages for cross-language clone detection. Additionally, we examined these models' potential for detecting single-language semantic clones. The models for this study were chosen based on their F1 scores\cite{lu2021codexglue} and availability to compare them with different datasets and techniques.

\section{Conclusion}
Code clone detection has widely been studied in software engineering research, and detecting semantic clones appears to be one of the most difficult tasks. Over the past few decades,  researchers have put extensive effort on designing techniques for detecting clones, as well as evaluating these techniques on available benchmark datasets. Despite the success in detecting syntactic clones, semantic clone detection and cross-language clone detection still appear to be a considerable challenge. To improve this context, it is critical to understand the relative performances of various clone detection models considering model and data aspects.

To improve our understanding of how various models compare when performing this task, we studied the cutting-edge models available for both single-language and cross-language clones. We found that single-language clone detection models' performances drop when tested with code fragments with less syntactical similarity. Although this is expected, we noticed the drops to have high variability across models indicating the inherent abilities of these models to detect semantic clones should be assessed more carefully. We also found that cross-language models tend to learn better representation in such scenarios. 

In summary, our evaluation demystifies the status of semantic clone detection, reveals the capabilities of cross-language clone detection models to be used for single-language clone detectors, and provides insights into how all these models perform under mutation operations. We believe that our evaluation technique would provide researchers with a pathway to evaluate future clone detection models and inspire them to explore the potential of language-agnostic clone detection models. 
\icsmepinku{
 \section{Acknowledgement}
 This work was supported by   NSERC Discovery, CFI-JELF, and NSERC CREATE graduate program on Software Analytics Research (SOAR) grants.
 }
 
% \balance
% \bibliographystyle{ieeetr}

\end{document}